\documentclass[journal=jctcce,manuscript=article]{achemso}

\usepackage{graphicx}
\graphicspath{ {figures_xyna/} }
\usepackage[outdir=./]{epstopdf}
\usepackage{subcaption}

\newcommand{\Conv}{\mathop{\scalebox{1.5}{\raisebox{-0.2ex}{$\ast$}}}} 

\usepackage{textcomp} 

\usepackage[T1]{fontenc} 

\author{Daniel J. Sharpe}
\author{Konstantin R\"{o}der}
\author{David J. Wales}
\email{dw34@cam.ac.uk}
\affiliation[University of Cambridge]
{Department of Chemistry, University of Cambridge, Lensfield Road,\\ Cambridge CB2 1EW, United Kingdom}

\title[Energy landscapes and dynamics of xylo-nucleic acids]
  {Energy landscapes and dynamics of xylo-nucleic acids}

\abbreviations{IR,NMR,UV}

\begin{document}

\begin{tocentry}

Some journals require a graphical entry for the Table of Contents.
This should be laid out ``print ready'' so that the sizing of the
text is correct.

Inside the \texttt{tocentry} environment, the font used is Helvetica
8\,pt, as required by \emph{Journal of the American Chemical
Society}.

The surrounding frame is 9\,cm by 3.5\,cm, which is the maximum
permitted for  \emph{Journal of the American Chemical Society}
graphical table of content entries. The box will not resize if the
content is too big: instead it will overflow the edge of the box.

This box and the associated title will always be printed on a
separate page at the end of the document.

\end{tocentry}

\pagebreak

\begin{abstract}
Artificial analogues of the natural nucleic acids have attracted recent interest as a diverse class of information storage molecules capable of self-replication. In the present study, we use the computational potential energy landscape framework to investigate the structural and dynamical properties of xylo- and deoxyxylo-nucleic acids (XyNA and dXyNA), which are derived from their respective RNA and DNA analogues by an inversion of configuration at a single chiral center in the sugar moiety of the nucleotide unit. The free energy landscapes of an octameric XyNA sequence and its dXyNA analogue demonstrate the existence of a facile conformational transition between a left-handed helix that is the global free energy minimum, and a closely competing ladder-type structure with approximately zero helicity. The separation of the competing conformational ensembles is better-defined for the dXyNA system, whereas the XyNA analogue is inherently more flexible. The former therefore appear more suitable candidates for a molecular switch. The landscapes differ qualitatively from those reported in previous studies for evolved biomolecules: they are significantly more frustrated, so that XyNAs provide an example of an unnatural system for which the conditions constituting the principle of minimal frustration are, as may be expected, violated.


\end{abstract}

\pagebreak

\section{Introduction}

Xeno-nucleic acids (XNAs) are a diverse family of nucleic acid structures derived from DNA or RNA by chemical modification of the nucleotide units.\cite{herdewijnbiodivers2009} XNAs have rapidly emerged as having important medical applications,\cite{wang2013} for example as aptamers,\cite{appellaopinion2009,tayloropinion2014} synthetic ribozymes,\cite{taylornature2015} artificial small interfering RNAs (siRNAs) and antisense oligonucleotides for the targeting of microRNAs.\cite{morihiro2017,pinheiroopinion2012} The resistance of XNAs to endonucleases, a result of the inability of natural enzymes to recognise the modified nucleic acid structures, is a particularly valuable advantage to the use of XNAs in place of natural nucleic acids for therapeutic purposes.\cite{chaputreview2012} XNAs are also of current interest in the emergent fields of synthetic biology,\cite{synthbiolreview} which demands the development of chemical information storage systems capable of self-replication \textit{in vitro} and \textit{in vivo} for artificial life and biological computation, and systems chemistry,\cite{systemschemreview} which requires molecular switches for the control of operations in complex chemical networks. Other potential applications of XNAs include their use as self-assembling nanomaterials, broadening the possible design scope in DNA nanotechnology, chemical sensors and catalysts.\cite{pinheirotrends2014,taylorchembio2016} The study of XNAs is also motivated by the fundamental question of the origins of life, where it is important to understand the factors that led to evolution selecting ribofuranosyl nucleic acids as the genetic biopolymer for the basis of life, and where it remains unknown if an alternative nucleic acid was utilised in hypothetical organisms preceding those based on RNA.\cite{taylornature2015} Despite their importance, there are few reported structures of XNAs, and relatively little is known concerning the structural and dynamical properties of XNAs in atomistic detail.\cite{anosovareview2016} In particular, computational studies have been thus far largely limited to molecular dynamics (MD) simulations, which do not overcome the broken ergodicity encountered in biomolecules.

\begin{figure}
\centering
\includegraphics[scale=0.2]{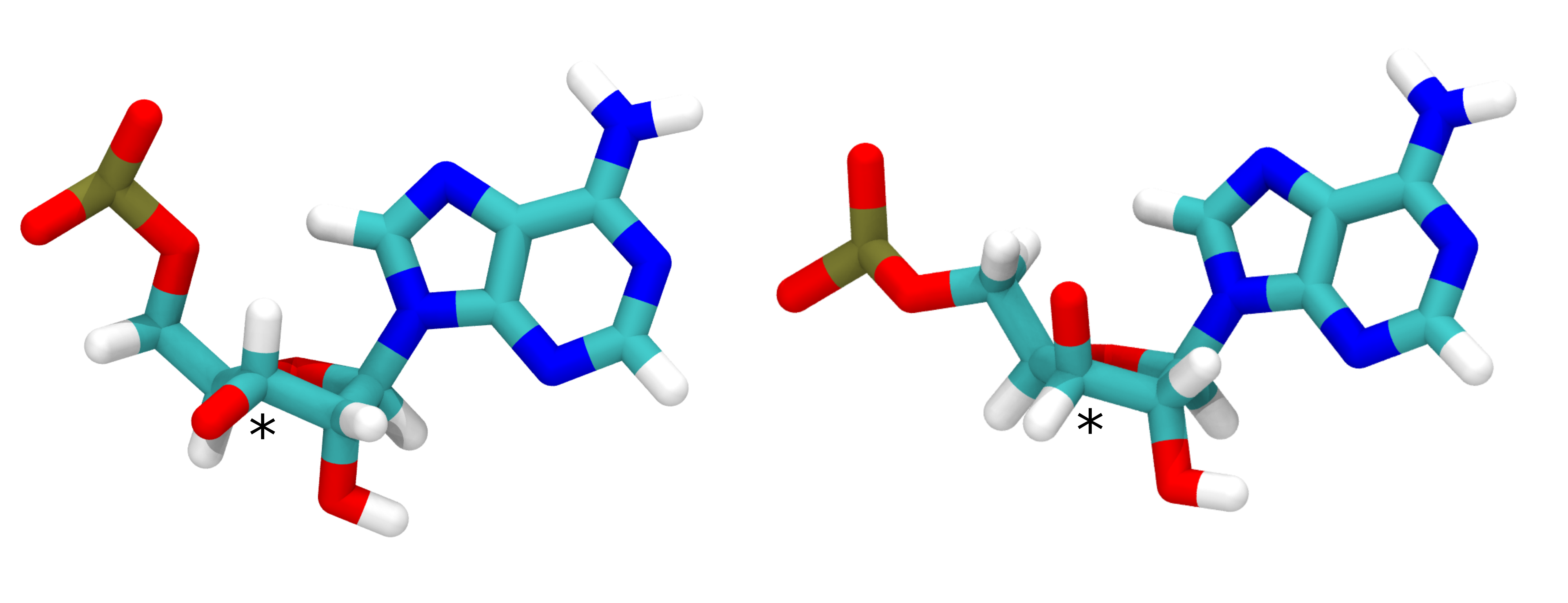}
\caption{Comparison of (left) a ribonucleotide monomer unit as it appears in canonical A-RNA, and (right) a xylonucleotide monomer unit as it appears in the ladder-type conformation of the XyNA1 sequence oberved by NMR. Xylose is derived from ribose by an inversion of configuration at the C3$'$ chiral centre, indicated by ($\protect\Conv$). Note that the sugar moieties in both units exist in the C3$'$-\textit{endo} conformation, so that the O3$'$ atom of XyNA units is axial, and that the glycosidic torsion angle differs between the two units.}
\label{fig:1}
\end{figure}

The present study is focused on nucleic acids based on xylose (XyNA) and deoxyxylose (dXyNA), herein referred to collectively as XyNAs, which represent some of the simplest possible perturbations to the chemical structure of natural nucleic acids. Xylose is derived from ribose by a simple inversion of configuration at the C3$'$ atom of the sugar moiety (Fig.\,\ref{fig:1}), and likewise deoxyxylose is derived from deoxyribose. Xylose is a thermodynamic product of the formose reaction,\cite{mullerhca1990} the most probable prebiotic route of sugar synthesis,\cite{orgelcritrev2004} and so XyNAs represent arguably the most credible possibility of a genetic biopolymer adopted by prebiotic organisms that are speculative precursors to RNA-based life forms.

Thermal denaturation studies have demonstrated that dXyNA:DNA hybrid duplexes exhibit markedly low thermodynamic stability compared to corresponding DNA duplexes,\cite{babu2005} whereas dXyNA homoduplexes display thermodynamic stability commensurate with analogous DNA duplexes.\cite{seela1996} In this respect, dXyNA exhibits complementary properties to many alternative XNAs, which are able to form a stable duplex through hybridisation with DNA or RNA, with strong discrimination against mismatches.\cite{anosovareview2016} While this behaviour precludes the use of XyNAs as an aptamer and for other applications requiring sequence-specific binding, it is an ideal property if XyNAs are to be utilised alongside and independent of natural nucleic acids as an orthogonal information system.

Circular dichroism (CD) studies have shown that XyNA and dXyNA oligomers may adopt a left-handed helical duplex structure or a structure with an apparent lack of helicity, and that the observed structure is dependent on base sequence and sequence length as well as external factors including temperature and salt concentration.\cite{schoppe1996,maiti2012}

MD simulations of XyNA and dXyNA duplexes of length 8, 13 and 29 bp (base pairs) have revealed the existence of a helical inversion transition from a right- to a left-handed helical conformation.\cite{ramaswamyjacs2010,ramaswamyjctc2017} The observed timescale for this transition is of the order of tens of ns. For XyNA duplexes, it was observed that the left-handed helical structure is not stable, but rather that the terminal regions of the duplex undergo oscillatory movements from coiled to uncoiled states that act to repeatedly screw and unscrew the helix, suggesting structural competition between left-handed helical and linear ladder-type structures. The existence of this equilibrium between a pair of competing conformational ensembles suggests XyNAs as a potential molecular switch, and is also a property unique to XyNAs among the known XNAs.\cite{anosovareview2016} Other recent MD studies of XyNA duplexes in the presence of a carbon nanotube have demonstrated fast spontaneous unzipping as a consequence of the strained backbone, highlighting the promise of XyNAs with respect to gene delivery for therapeutic purposes.\cite{ghoshjpcc2016,ghoshjpcb2016}

In the present study, we use the computational potential energy landscape framework\cite{walesannurevphyschem2018} to investigate the structural and dynamical properties of XyNA and dXyNA duplexes. The extensive sampling facilitated by discrete path sampling\cite{walesphiltrans2012} (DPS) allows for the calculation of free energies and therefore proper comparison of the relative thermodynamic stabilities of the three major conformations expected, namely left-handed helical, right-handed helical and ladder-type structures. The free energy barriers partitioning these major conformational ensembles determine the applicability of XyNA and dXyNA duplexes as molecular switches, which requires two competing funnels to be separated by a barrier that is surmountable at ambient temperatures. Visualisation of the free energy landscapes will clearly elucidate structural differences between XyNA and dXyNA duplexes, as well as between them and their naturally evolved counterparts. The framework has been successfully applied to a variety of biomolecular systems\cite{biomolecularlandscapereview} including the B-Z-DNA transition\cite{chakrabortypccp2017} and the formation of DNA mini-dumbbells.\cite{klimaviczjctc2018}

\section{Methodology}

\subsection{System modeling}

The xylo- and deoxyxylo-nucleotide monomer units were constructed with the furanose moiety in the C3$'$-\textit{endo} (\textit{cf}. canonical A-RNA) and C2$'$-\textit{endo} conformations, respectively, using the LEAP program of AMBER.\cite{amber12} Atom parameters were given by the parm99 force field\cite{parm99} incorporating the bsc0 correction\cite{amberbsc0} for $\alpha$ and $\gamma$ backbone torsion angles. The parm99 force field has been applied to yield important insight into the behaviour of a variety of artificial nucleic acid systems,\cite{ivanonvajpca2007,froeyenbmc2016,veronanature2017} and can therefore be justifiably applied to XyNAs as in previous MD studies.\cite{ramaswamyjacs2010,ramaswamyjctc2017} The bsc0 correction is acknowledged to yield a general improvement in the description of the behaviour of both DNA and RNA, including noncanonical structures thereof, hence its implementation in the present work.
It was previously shown that different parameterisations for the glycosidic torsion angle $\chi$ have little effect on the MD trajectories of XyNA sequences,\cite{ramaswamyjctc2017} therefore no reparameterisation of this dihedral was used.
Partial charges were obtained by the two-stage RESP fitting procedure\cite{amberresp} at the HF/6-31G* level of theory using the ANTECHAMBER program\cite{antechamber} of AMBER, with electrostatic potentials calculated using Gaussian03.\cite{gaussian03} The potential function was correctly symmetrised.\cite{ambersymm2010,ambersymm2010corr}

With the nucleotide units constructed as described, complete right-handed duplexes of the octameric sequence (5$'$-3$'$)[xG-xU-xG-xU-xA-xC-xA-xC-T] (XyNA1) and the deoxyxylose analogue thereof (dXyNA1) were constructed using the LEAP program based on a template of canonical B-DNA produced with the NAB program\cite{ambernab} of AMBER. Ladder-type structures were obtained from the NMR solution structure of the XyNA oligomer (PDB: 2N4J).\cite{2n4jpdb} From these initial structures, optimised right-handed helical and ladder-type structures were obtained by basin-hopping global optimisation,\cite{doyeprl1998} as described below. From the latter conformation, in turn, a left-handed helical structure was obtained by short timescale explicit solvent MD simulation, performed using the SANDER package\cite{caseamber} of AMBER14, and further optimised by basin-hopping global optimisation.

\subsection{Exploration of the energy landscapes}

Basin-hopping (BH) global optimisation\cite{walesjpca1997,doyeprl1998,doyejcp1998,lipnas1987,lijmolstruct1988} to obtain the lowest-energy structures of the right-handed helical, left-handed helical and ladder-type conformational ensembles was achieved using the GMIN program\cite{GMIN} interfaced with AMBER12. The BH algorithm uses a Metropolis criterion to sample basins of attraction of a potential energy landscape and thereby locate low-energy configurations. Perturbation moves include concerted rotations of groups about internal axes of the system, in addition to standard atomic displacements. Chirality checks are implemented to ensure that no chiral centres become inverted in the course of the simulation.

The landscapes were further explored using discrete path sampling (DPS) to create kinetic transition networks (KTNs).\cite{Wales2003,walesmolphys2002,walesmolphys2004,walescurropstructbiol2010,noecurropinstructbiol2008} Transition states were located with the doubly-nudged elastic band algorithm\cite{neb4,neb1,neb2,dneb,dnebcorrection,sheppardjcp2008} and converged with hybrid eigenvector-following.\cite{walesjcp1994_ef,walesjcp1996_ef,maurojpca2005,hef,mantelljctc2016} Local minima were then characterised by approximate steepest-descent paths, using a modified version of the L-BFGS algorithm.\cite{nocedalbook,lbfgs} These calculations used the OPTIM program\cite{OPTIM} interfaced with AMBER12. Initial optimal alignment of endpoint structures was achieved by a shortest augmenting path alogorithm.\cite{walesqci} A Dijkstra-based missing connection algorithm\cite{dijkstra,carrjcp2005-2} was used to construct a priority list of connection attempts based on appropriate edge weights.

After initial sampling initiated from the low energy structures located by BH, the PATHSAMPLE driver program\cite{PATHSAMPLE} was used to conduct further sampling to improve the connectivity of the landscape,\cite{roederjacs2018} remove artificial kinetic traps and high energy barriers,\cite{strodeljacs2007} and shorten path lengths.\cite{carrjcp2005}

All calculations described above used a generalised Born implicit solvent model with surface area term\cite{onufrievproteins2004,onufrievjpcb2000} and an effective monovalent salt concentration of 0.1 M maintained using the Debye-H\"{u}ckel approximation.\cite{srinivasantca1999}

Free energy landscapes at a temperature of 298 K were calculated using the harmonic superposition approximation (HSA).\cite{walesmolphys1993,doyejcp1995,strodelcpl2008} A self-consistent recursive regrouping scheme based on a specified free energy barrier threshold was used to classify minima into collective macrostates.\cite{carrjpcb2008,walespnas2014}
Free energy pathways were determined by Dijkstra's shortest path algorithm.\cite{dijkstra,evansjcp2004}

\subsection{Analysis of the energy landscapes}

Free energy landscapes are visualised as disconnectivity graphs,\cite{beckerjcp1997,walesnature1998,krivovjcp2002,evansjcp2003,roederjacs2018} where each leaf corresponds to a free energy group, and is coloured according to the value of an appropriate order parameter for a representative potential energy minimum of that group. Here, the chosen order parameter is the helical handedness $H$,\cite{saguipnas2009} in the form originally proposed for describing the B-Z transition in DNA duplexes.\cite{saguinar2013} The magnitude of $H$ is a measure of the number of turns associated with a helix. Values $H > 0$ and $H < 0$ are reflective of right- and left-handed double helial turns, respectively, while values $H \approx 0$ indicate approximately zero helicity.

Processing of the data in the KTN was achieved with the CPPTRAJ module\cite{cpptraj} of AMBER. The CURVES+ program\cite{curvesplus} was used to extract bp-axis, inter-bp and intra-bp geometric parameters.\cite{luolsonjmolbiol1999} All analyses exclude the terminal base pairs to avoid the influence of end effects. Molecular graphics images were produced using VMD.\cite{vmd}

\section{Results}

\subsection{Free energy landscapes}

\begin{figure}
\centering
\includegraphics[scale=0.7]{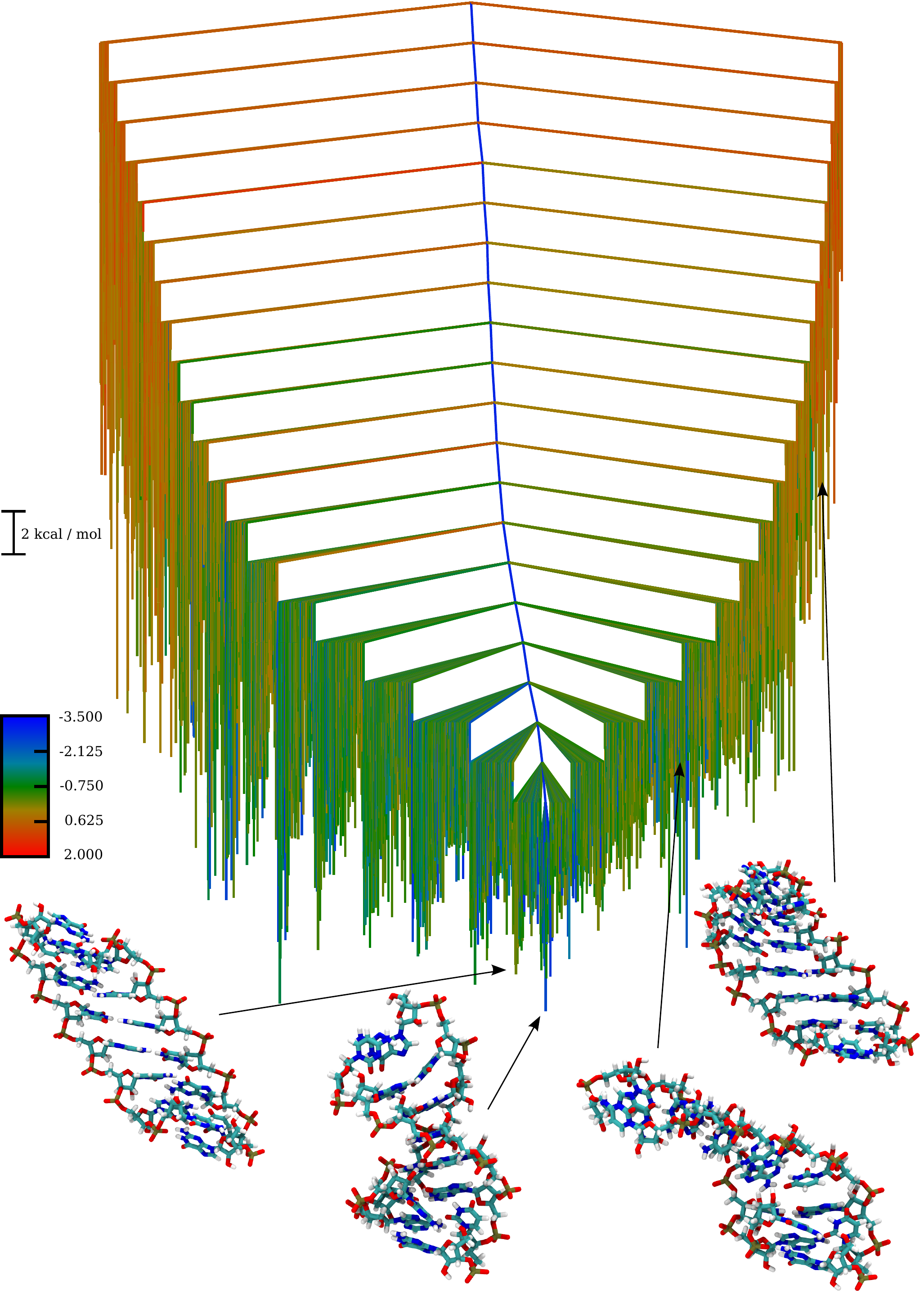}
\caption{Free energy landscape for the XyNA1 duplex at a temperature of 298\,K, using a regrouping threshold of 4\,kcal\,mol$^{-1}$ and a disconnectivity threshold increment of 2\,kcal\,mol$^{-1}$. The branches are coloured according to the helical handedness ($H$) of a single potential energy minimum representative of the free energy group. Some important representative structures from the different conformational ensembles are shown.}
\label{fig:2}
\end{figure}

\begin{figure}
\centering
\includegraphics[scale=0.7]{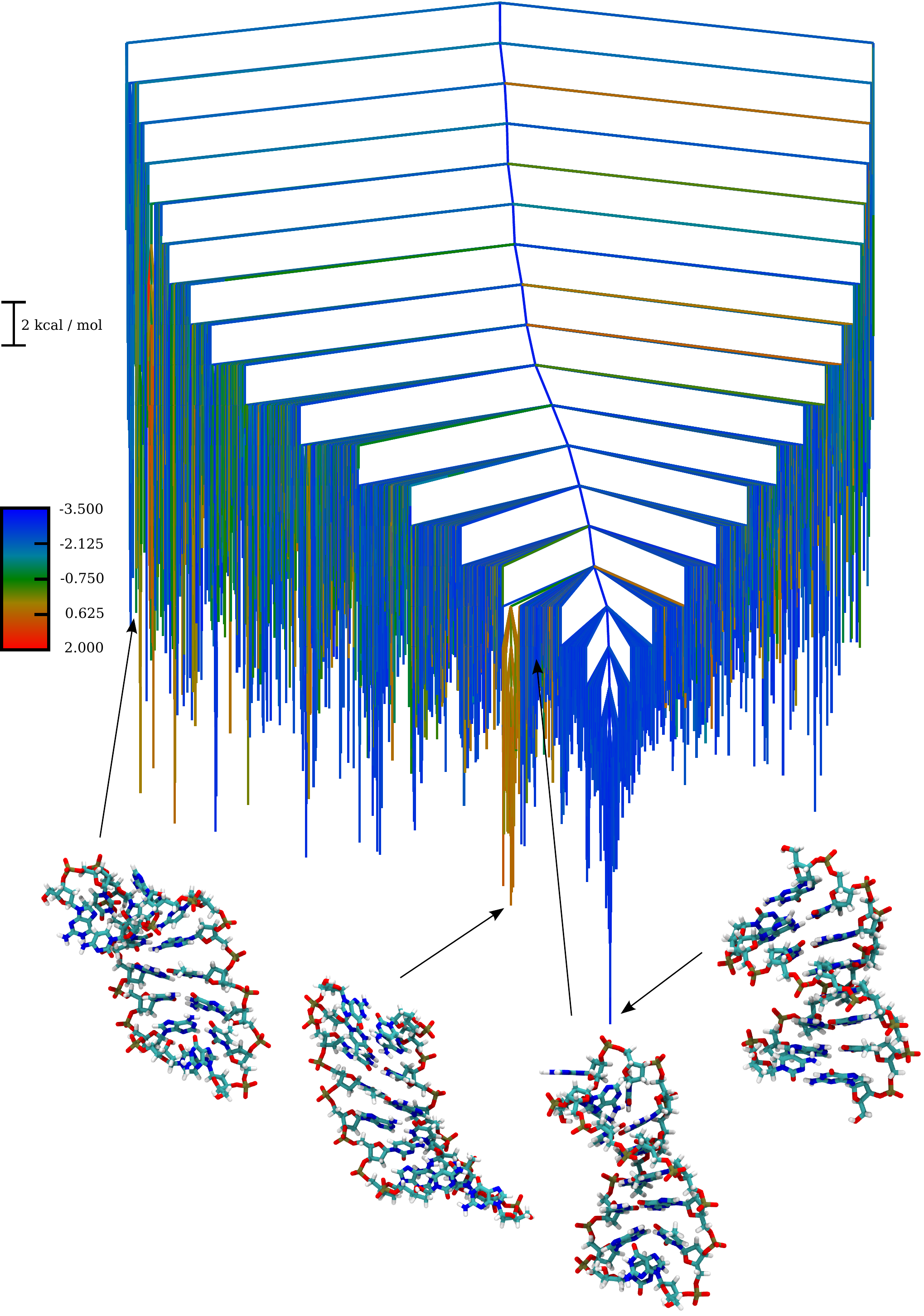}
\caption{Free energy landscape for the XyNA1 duplex at a temperature of 298\,K, using a regrouping threshold of 4\,kcal\,mol$^{-1}$ and a disconnectivity threshold increment of 2\,kcal\,mol$^{-1}$. The colour scale for the helical handedness ($H$) is the same as for Fig. \ref{fig:2}. Some important representative structures from the different conformational ensembles are shown.}
\label{fig:3}
\end{figure}

The free energy disconnectivity graphs at 298 K for XyNA1 and dXyNA1 are shown in Figs.\,\ref{fig:2} and \ref{fig:3}, respectively. For both duplexes, the right-handed helical structures are located only in regions of the landscape with high free energy, so that the occupation probability for right-handed helices is very low. The inversion of chirality in the nucleotide units, with respect to natural nucleic acids, seeds an inversion of preferred helical sense to favour a stable left-handed helix, and XyNAs can be thought of as effective two-state systems, with equilibrium between left-handed helical and ladder-type structures.

There are striking differences between the free energy landscapes for XyNA and dXyNA duplexes. Most notably, the ideal left-handed helical state of the dXyNA1 duplex is a relatively well-defined global free energy minimum on the landscape, being \textit{ca}. 5\,kcal\,mol$^{-1}$ more stable than the next lowest energy minimum, which is the most stable ladder-type structure. The free energy barrier for the conversion of the left-handed helical to the ladder-type structure is around 20\,kcal\,mol$^{-1}$, and for the reverse transition is around 12\,kcal\,mol$^{-1}$. By contrast, the free energy barriers separating the corresponding states of the XyNA1 duplex are around 10\,kcal\,mol$^{-1}$ in either direction, and their free energy difference is less than 2\,kcal\,mol$^{-1}$. Another immediately apparent difference between the free energy landscapes of the two systems is in the distribution of values for the helical handedness order parameter. For the XyNA duplex, a much broader range of values for $H$ is observed, with values ranging from $H \approx 0$ to $H \approx -3.0$ appearing in the low-energy region of the landscape. 
In notable contrast, the low free energy region of the landscape for the dXyNA duplex is dominated by left-handed helical structures, the handedness of which takes a somewhat narrow distribution of values \textit{ca}. $H \approx -3.0$. Thus XyNA duplexes are more flexible than their dXyNA analogues. The clear separation of free energy funnels for left-handed helical and ladder-type conformational ensembles in dXyNA makes it a more suitable candidate for use as a molecular switch than XyNA.

\begin{figure}
\centering
\includegraphics[scale=0.2]{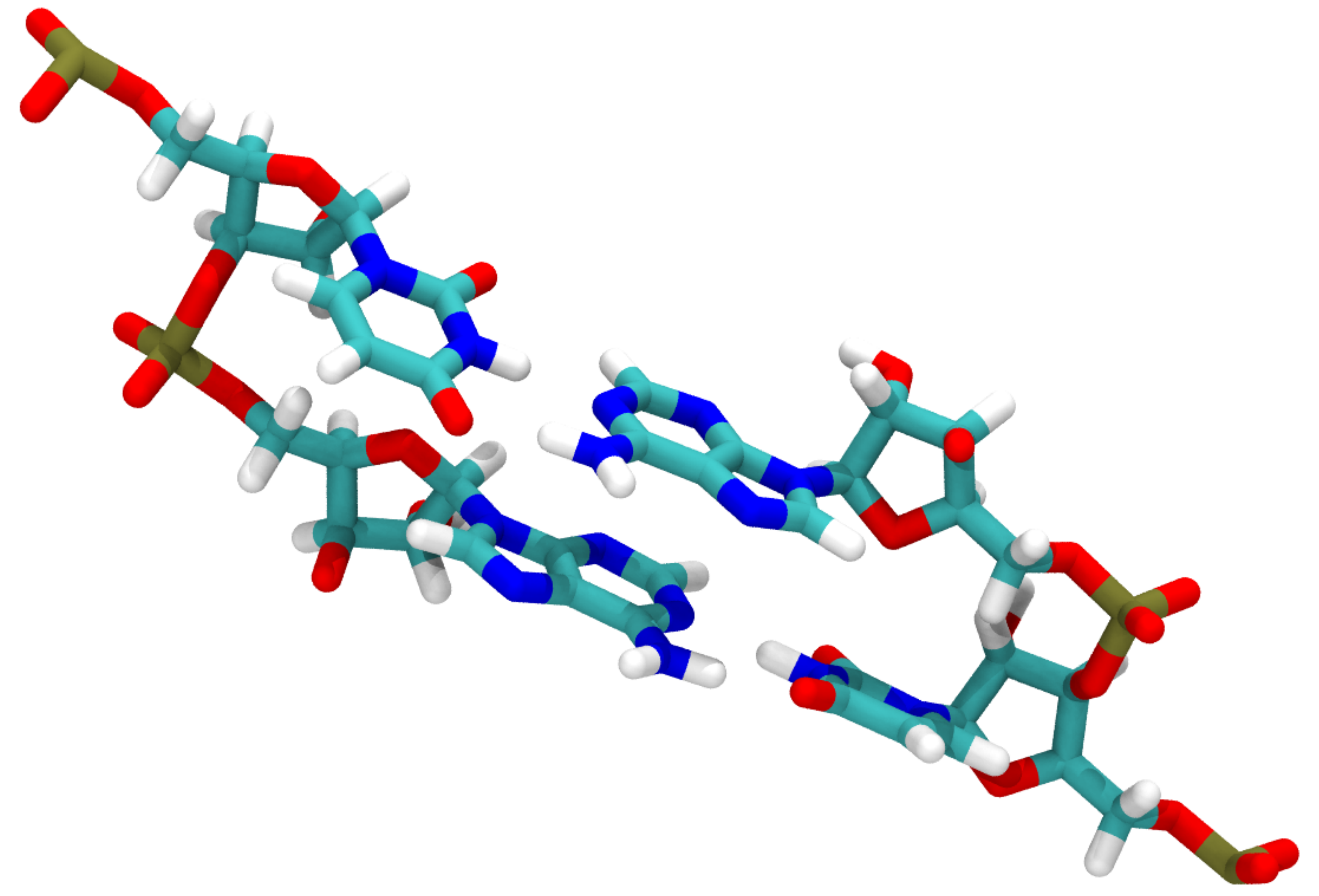}
\caption{A dinucleotide step in a structure that is part of the lowest free energy group in the conformational ensemble of ladder-type structures for XyNA1. The favourable interstrand stacking of adjacent bases is apparent.}
\label{fig:7}
\end{figure}

The ideal ladder-type structure of the XyNA1 duplex compares favourably with the NMR solution structure.\cite{2n4jpdb}
As noted in Ref.\,\citenum{maiti2012}, the ladder-type structures of the XyNA and dXyNA duplexes are stabilised by the interstrand stacking of adjacent bases, which arises due to the strong inclination of bases with respect to the helical axis (Fig.\,\ref{fig:7}).
However, the ladder-type structure is predicted to have marginal left-handed helicity, as opposed to the marginal right-handed helicity in the observed structure. The over-stabilisation of left-handed helical states of XyNA within the force field parameterisation implemented in this work is also evidenced by the fact that left-handed helical structure is erroneously predicted to be the global free energy minimum, although the free energy difference from the idealised ladder-type structure is very small. The force field is apparently more accurate in replicating the behaviour of dXyNA duplexes, where there appears to be no such bias, and the ladder-type structures are correctly predicted to have slight right-handed helicity.

The landscapes of both XyNA1 and dXyNA1 duplexes are significantly frustrated, there being many minima of low free energy in competition with the global free energy minimum. This finding is in contrast to the free energy landscape reported in a study of the analogous B-Z transition in CG-rich DNA sequences, where there is strong funnelling to the native B-DNA state. Frustration in the free energy landscapes of XyNAs is likely largely attributable to the geometrical frustration of base pairs that prevents formation of a properly extended linear duplex and instead necessitates that the ladder-type conformation of XyNA duplexes partly bends inwards on itself to maintain optimal Watson-Crick base-pairing of all base pairs, and that drives the continual unwinding and rewinding of the left-handed helix. This geometric frustration may also be relieved by the adoption of noncanonical base pairings at one or both of the duplex termini, which then allows the linear structures to form a properly extended conformation.

Evolved biomolecules obey the principle of minimal frustration.\cite{tzulpnas2017} The evolutionary process favours energy landscapes having strong bias towards a well-defined native state, and promotes the elimination of kinetic traps on the pathways to this native state. The fact that the energy landscapes of XyNAs are significantly more frustrated than the landscapes of DNA and RNA presents an evolutionary argument for the adoption of DNA and RNA over XyNAs.

\subsection{Free energy pathways}

\begin{figure}
\centering
\begin{subfigure}[b]{0.7\textwidth}
\includegraphics[width=\textwidth]{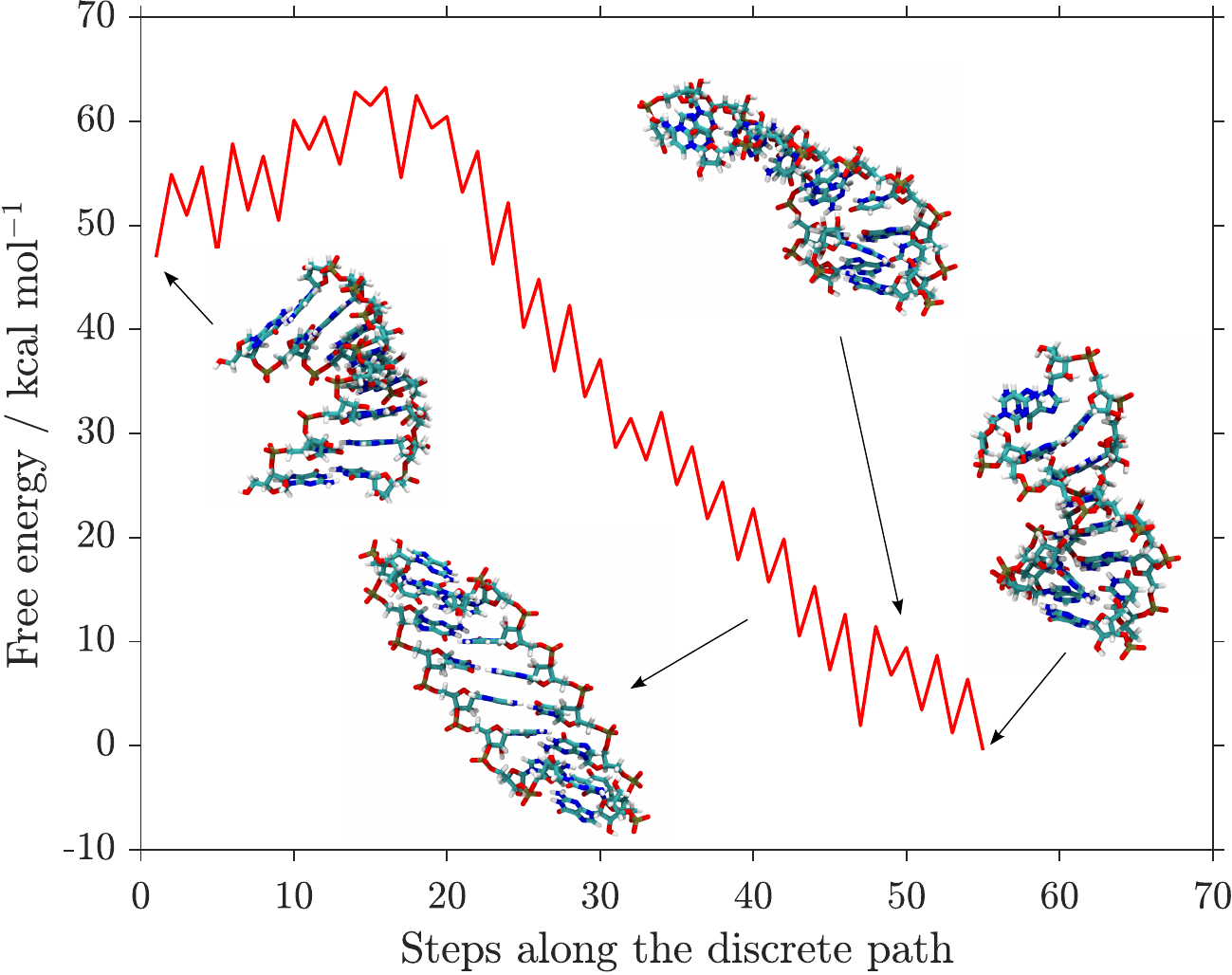}
\caption{Fastest pathway energy profile for the helical inversion transition of the XyNA1 duplex.}
\label{fig:10}
\end{subfigure}
\begin{subfigure}[b]{0.7\textwidth}
\includegraphics[width=\textwidth]{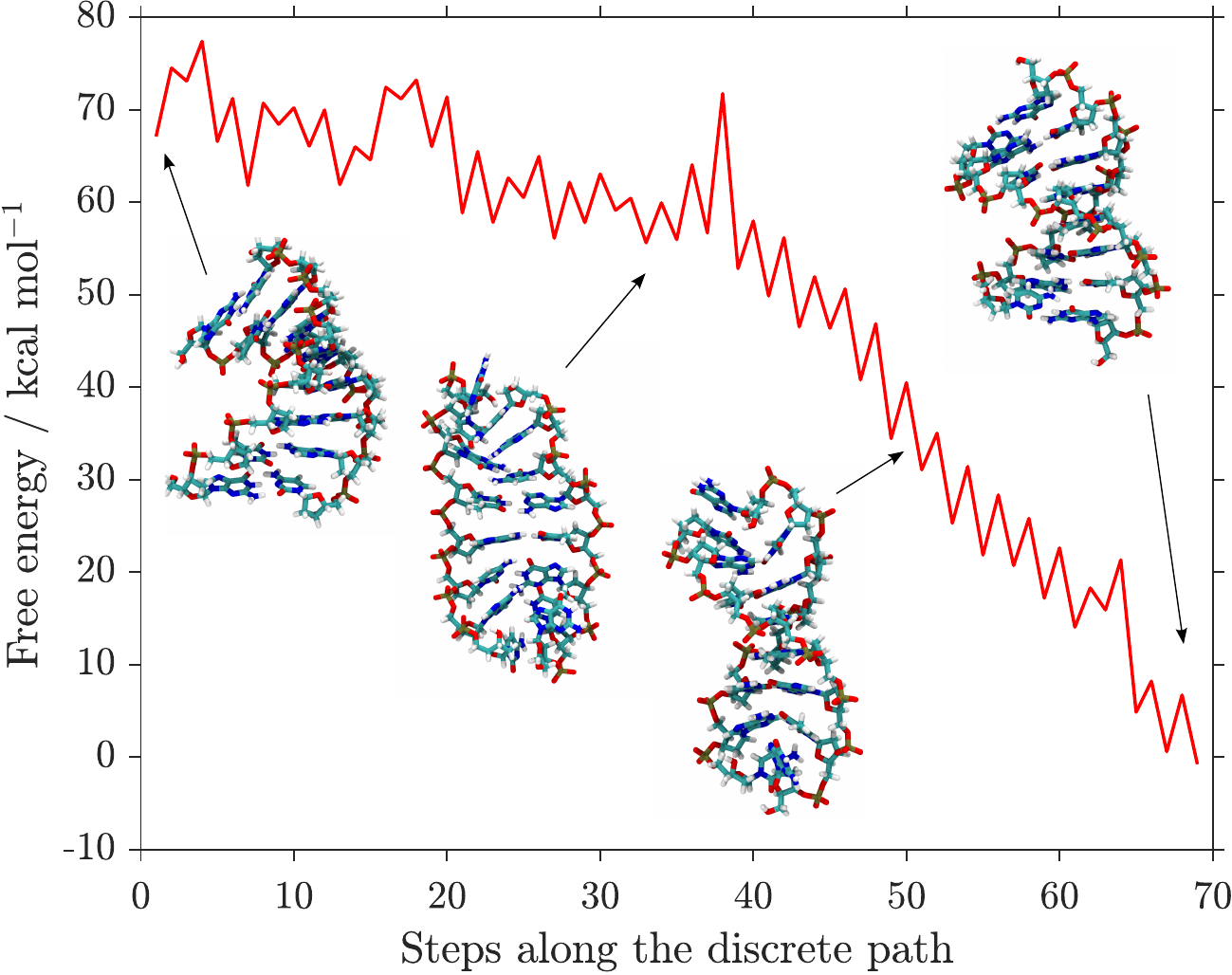}
\caption{Fastest pathway energy profile for the helical inversion transition of the dXyNA1 duplex.}
\label{fig:11}
\end{subfigure}
\caption{Free energy pathways for the right- to left-handed helical transitions in XyNA and dXyNA duplexes that make the single largest contribution to the steady-state rate constants.
Some representative structures that are key intermediates or transition states are included.}
\label{fig:4}
\end{figure}

Free energy pathways for helical inversion are shown in Fig.\,\ref{fig:4} for the XyNA1 and dXyNA1 sequences, going from a representative right-handed helical structure to a left-handed helix representing the global free energy minimum. For both duplexes, the pathways are downhill in energy, and the mechanism features only low barriers, leading to fast kinetics. Furthermore, the interconversion of ladder-type and left-handed helical states occurs by helix winding (unwinding) in left- (right-) handed directions, respectively, propagated inwards from one terminus, and not from both termini simultaneously.
[Calculate NGT rate consts (for ladder-type - left-handed helical transition), a key experimental observable]

For the XyNA1 duplex, the helical inversion proceeds via a low-energy ladder-type structure, which subsequently evolves to the left-handed helical state via a transition-state ensemble of `kinked' structures. The corresponding pathway for the dXyNA1 duplex is significantly different. In particular, structures with approximately zero helicity represent a high-energy transient state in the early stages of the pathway, and so the mechanism is not mediated by a ladder-type intermediate state, as is observed for the XyNA1 duplex. This again reflects the greater bias towards left-handed helical over ladder-type structures in dXyNA compared to XyNA duplexes. The transition then continues to progress smoothly with respect to handedness, that is, via a more regular helical structure with partial left-handed helicity.

\begin{figure}
\centering

\begin{subfigure}[b]{0.38\textwidth}
\includegraphics[width=\textwidth]{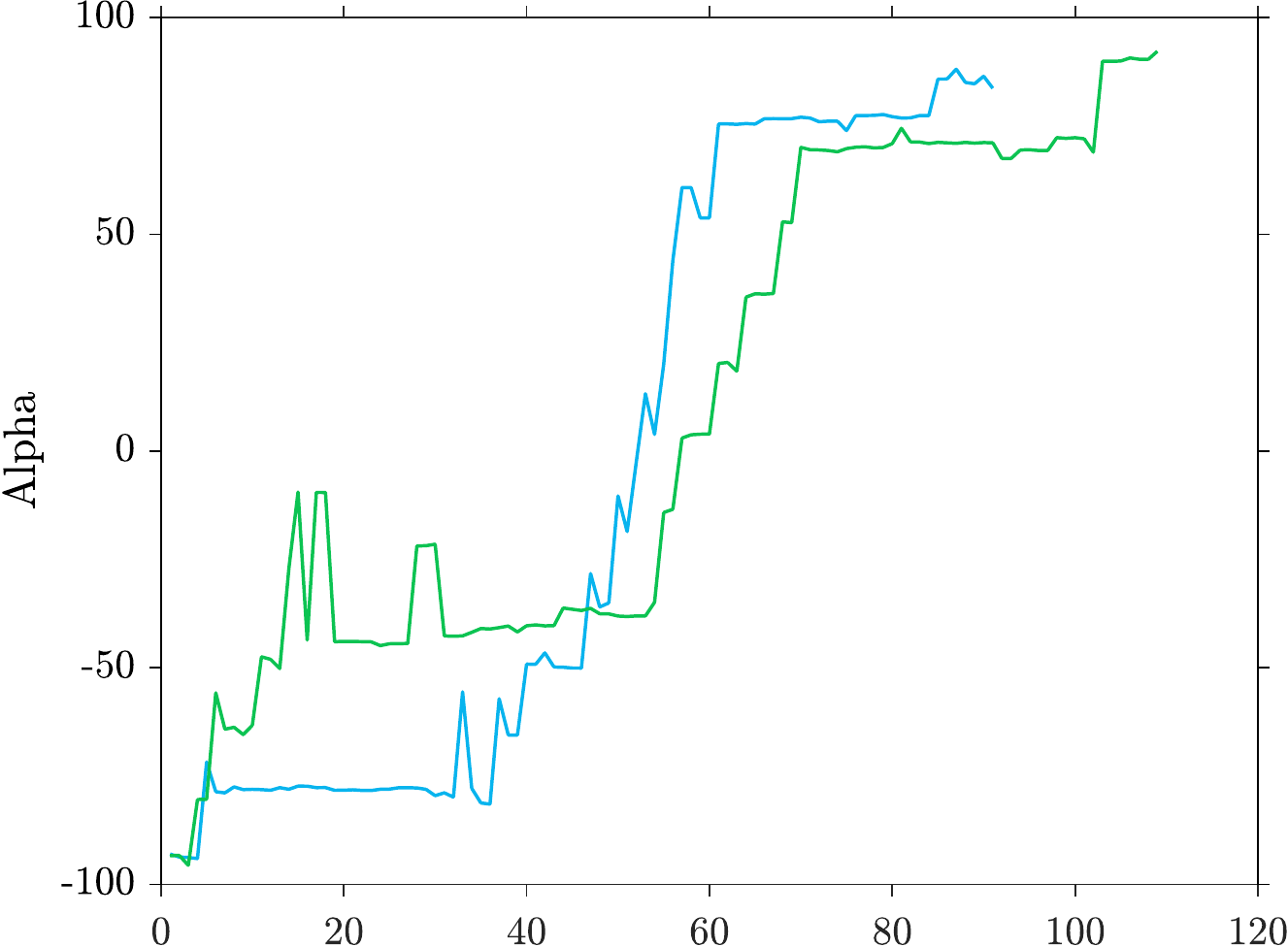}
\label{fig:19}
\end{subfigure}
\begin{subfigure}[b]{0.38\textwidth}
\includegraphics[width=\textwidth]{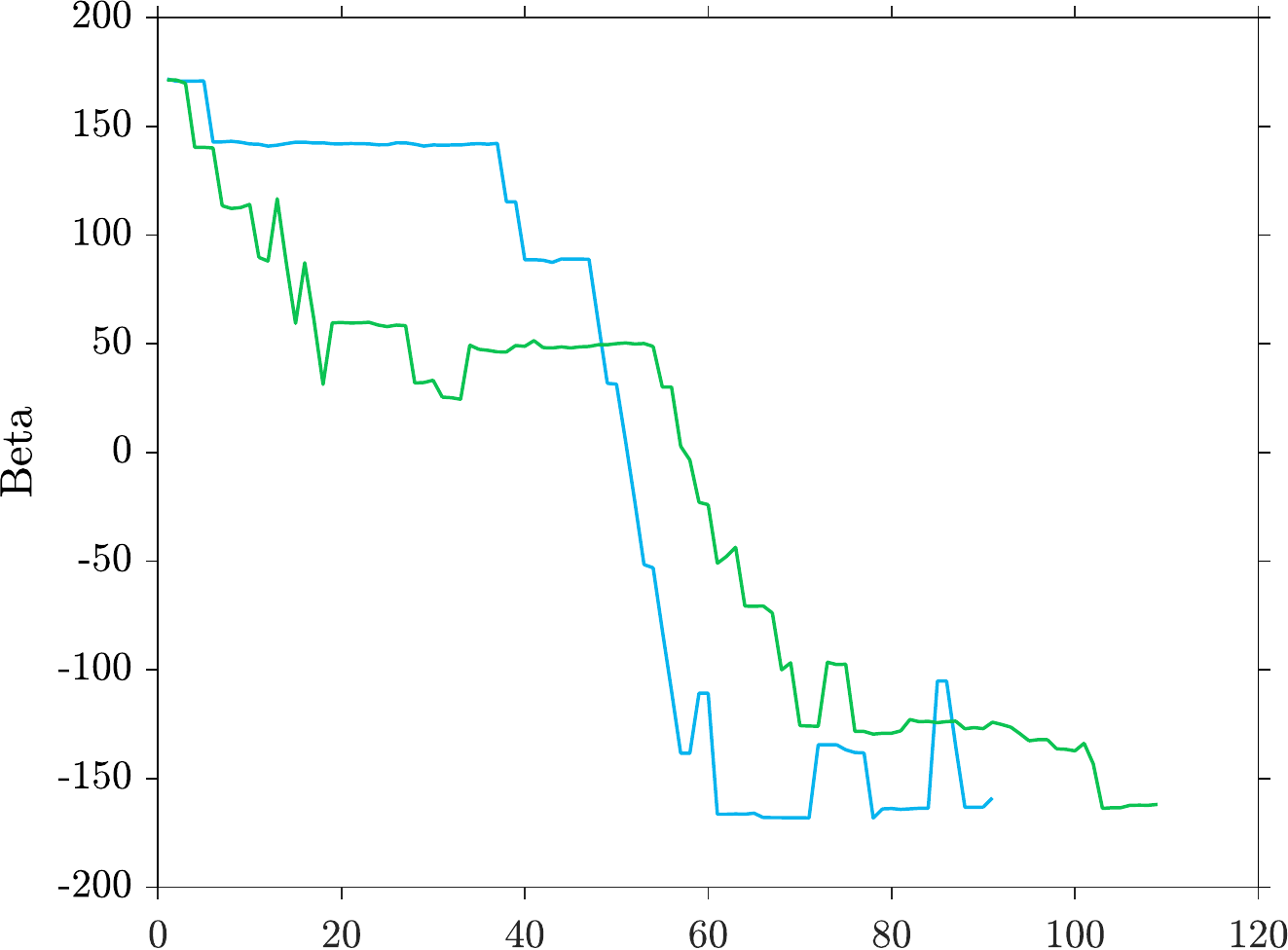}
\label{fig:20}
\end{subfigure}
\begin{subfigure}[b]{0.38\textwidth}
\includegraphics[width=\textwidth]{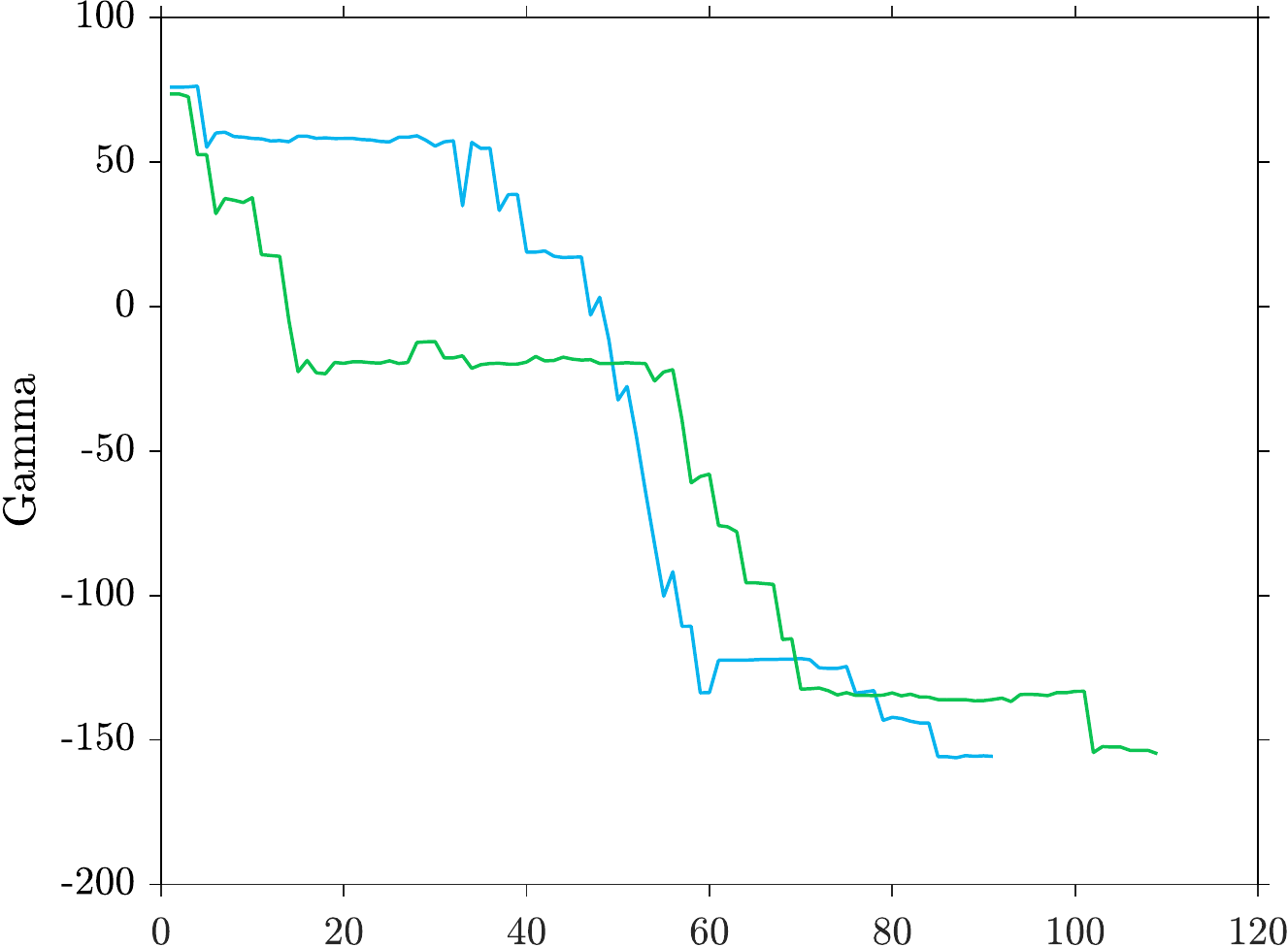}
\label{fig:21}
\end{subfigure}
\begin{subfigure}[b]{0.38\textwidth}
\includegraphics[width=\textwidth]{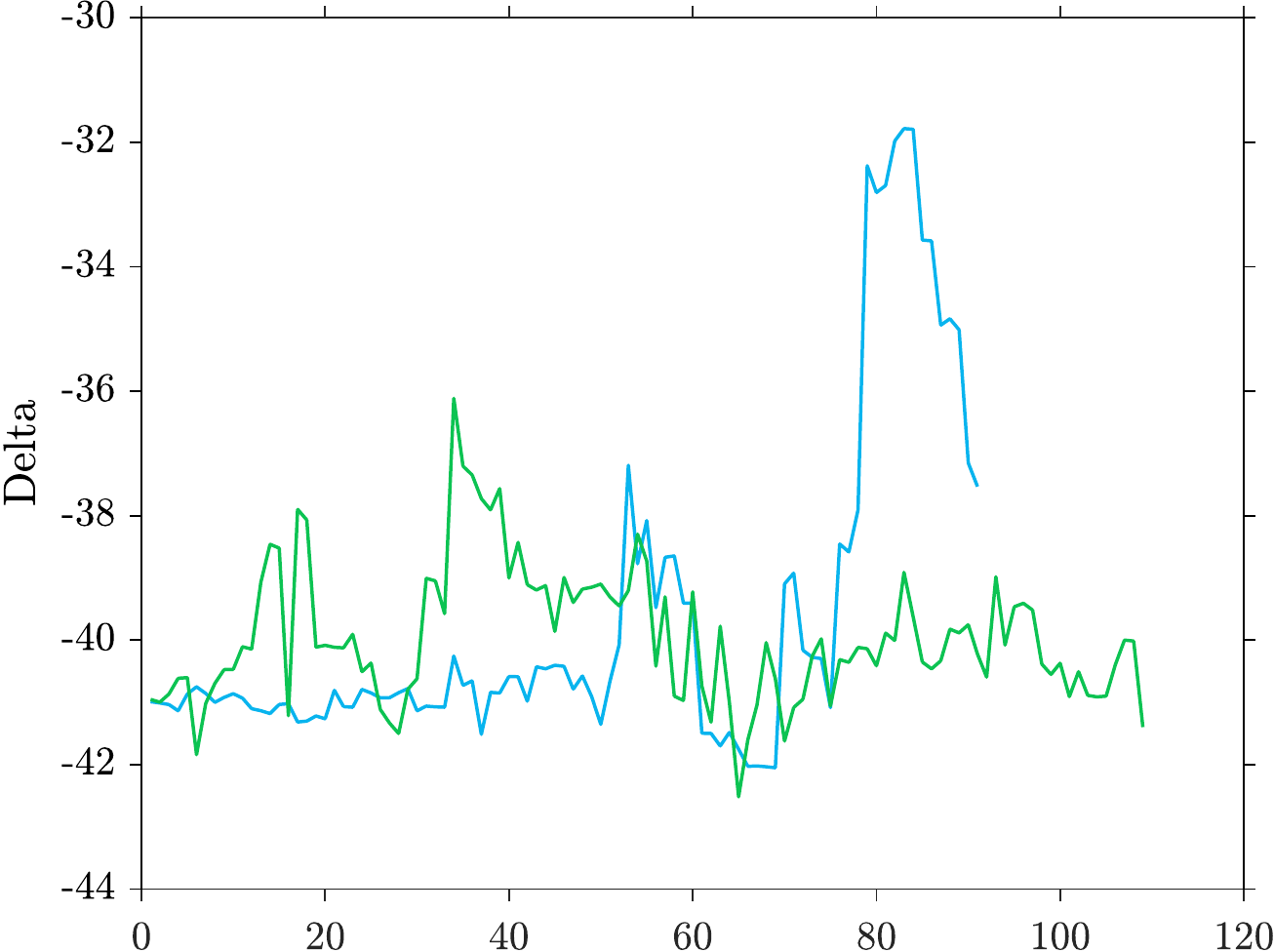}
\label{fig:22}
\end{subfigure}
\begin{subfigure}[b]{0.38\textwidth}
\includegraphics[width=\textwidth]{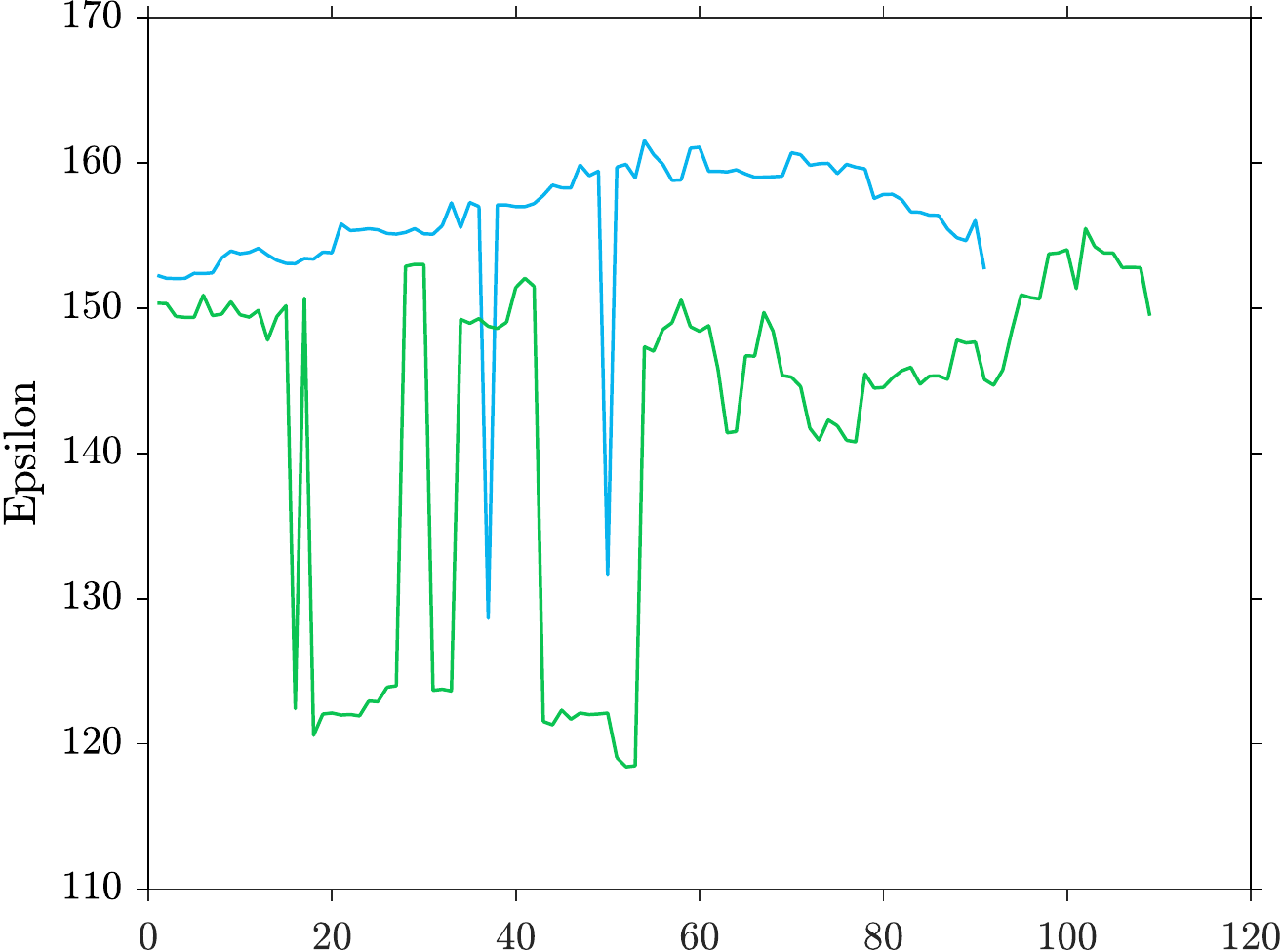}
\label{fig:23}
\end{subfigure}
\begin{subfigure}[b]{0.38\textwidth}
\includegraphics[width=\textwidth]{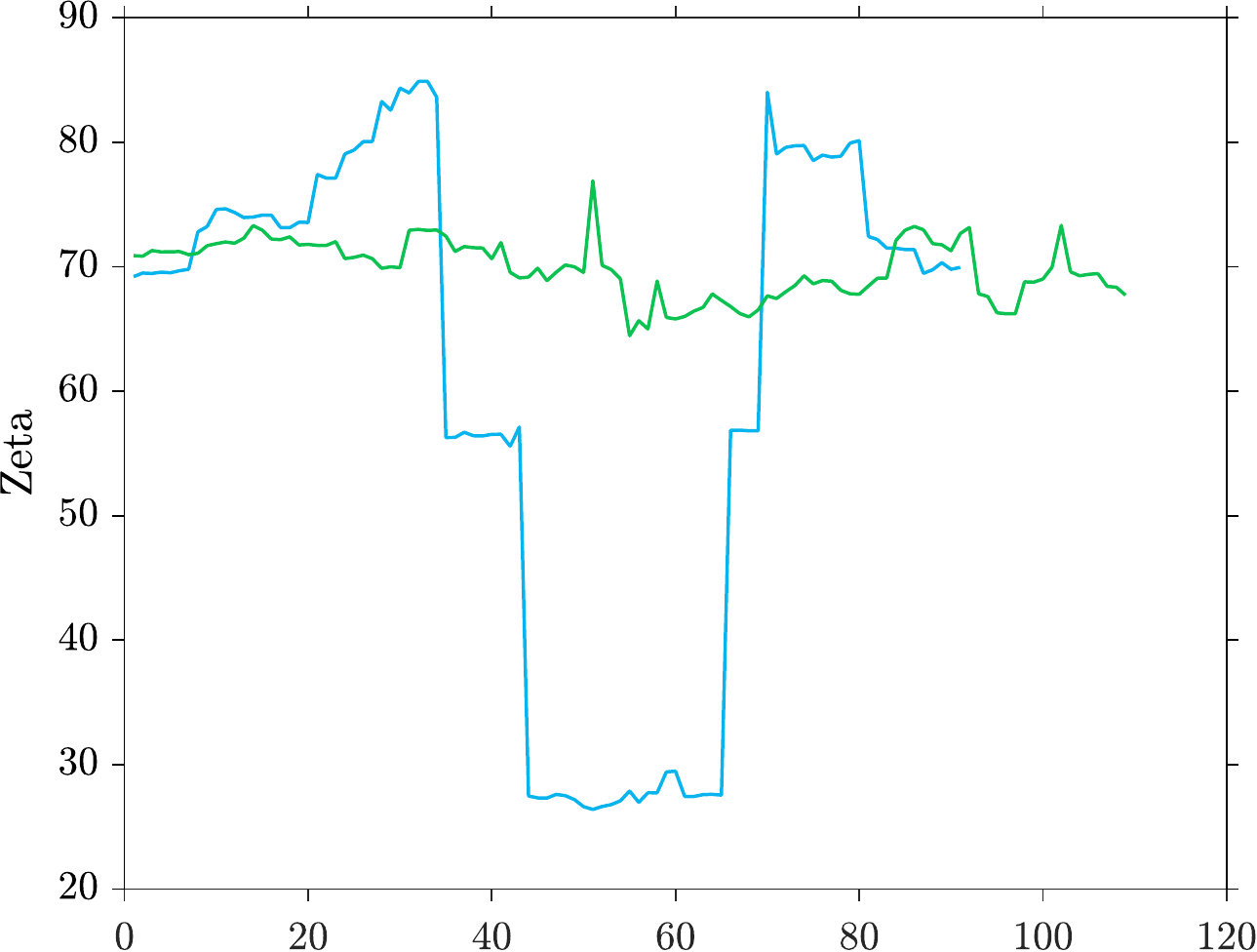}
\label{fig:24}
\end{subfigure}
\begin{subfigure}[b]{0.42\textwidth}
\includegraphics[width=\textwidth]{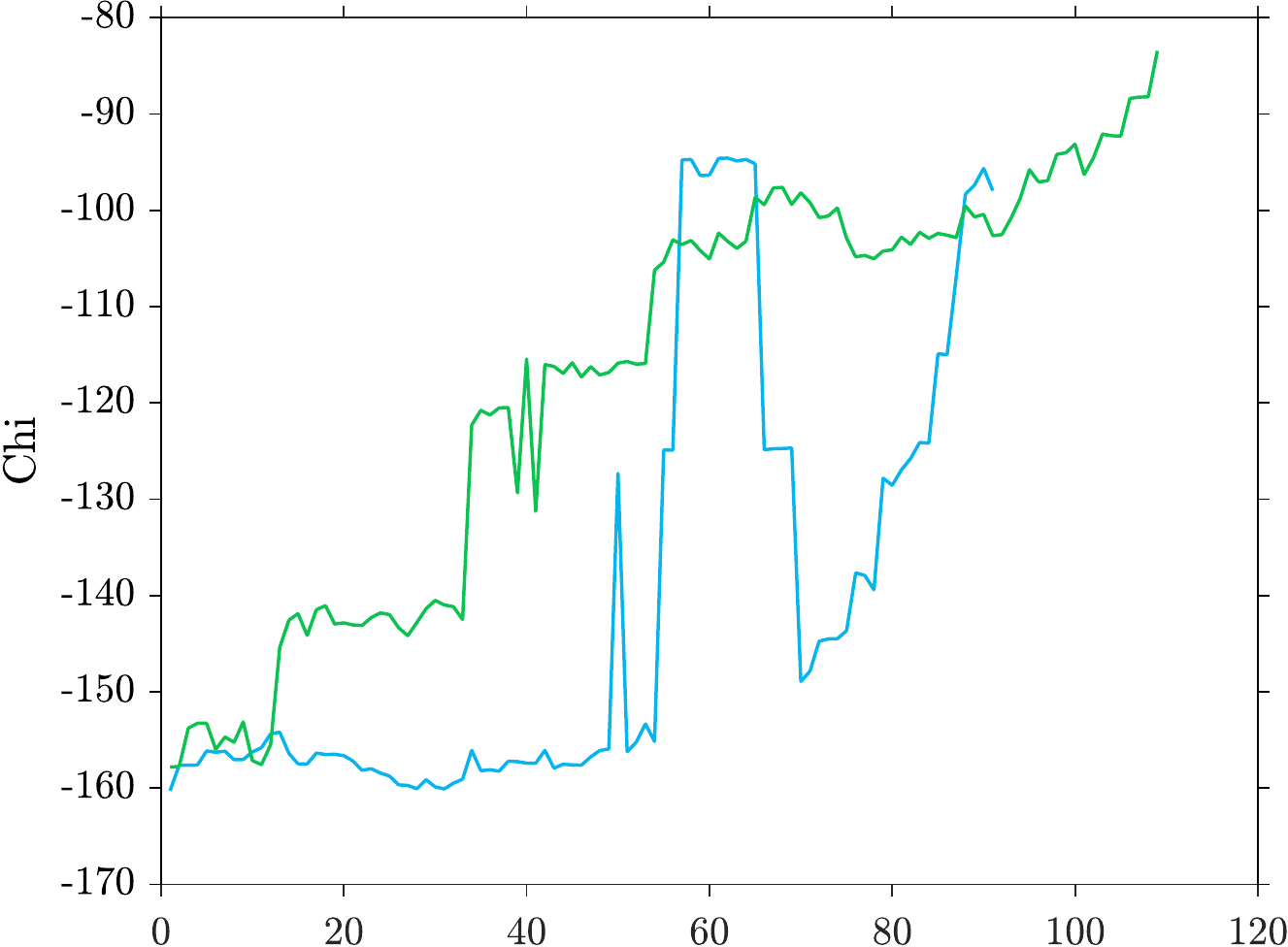}
\label{fig:25}
\end{subfigure}
\caption{Evolution of the backbone dihedral angles and glycosidic torsion angle $\chi$ (in deg.) along the fastest potential energy pathways for the right- to left-handed helical transition of the XyNA1 (blue) and dXyNA1 (green) duplexes, as a function of the position of the minima in the discrete path.}
\label{fig:5}
\end{figure}

\begin{figure}
\centering

\begin{subfigure}[b]{0.4\textwidth}
\includegraphics[width=\textwidth]{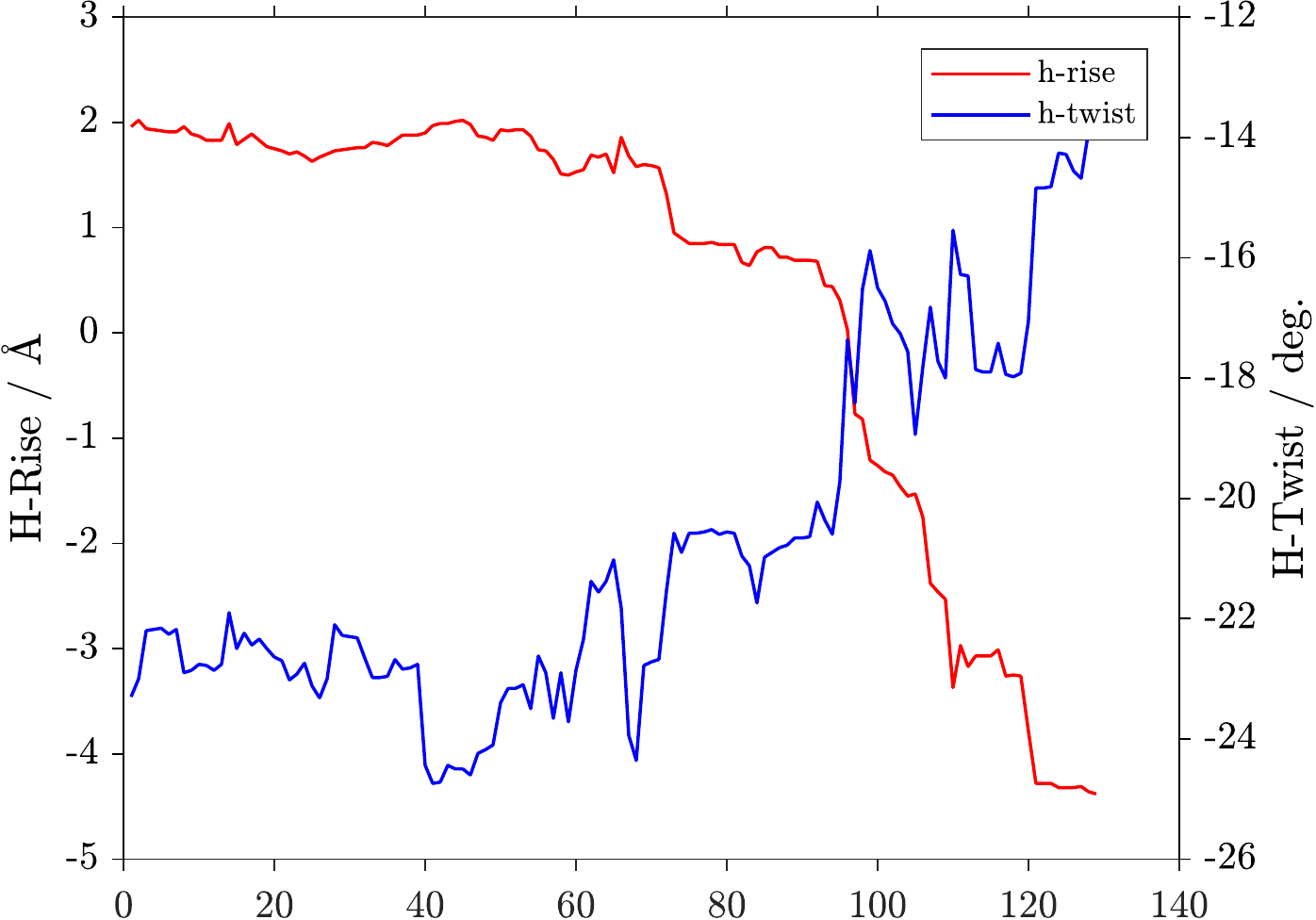}
\label{fig:61}
\end{subfigure}
\begin{subfigure}[b]{0.4\textwidth}
\includegraphics[width=\textwidth]{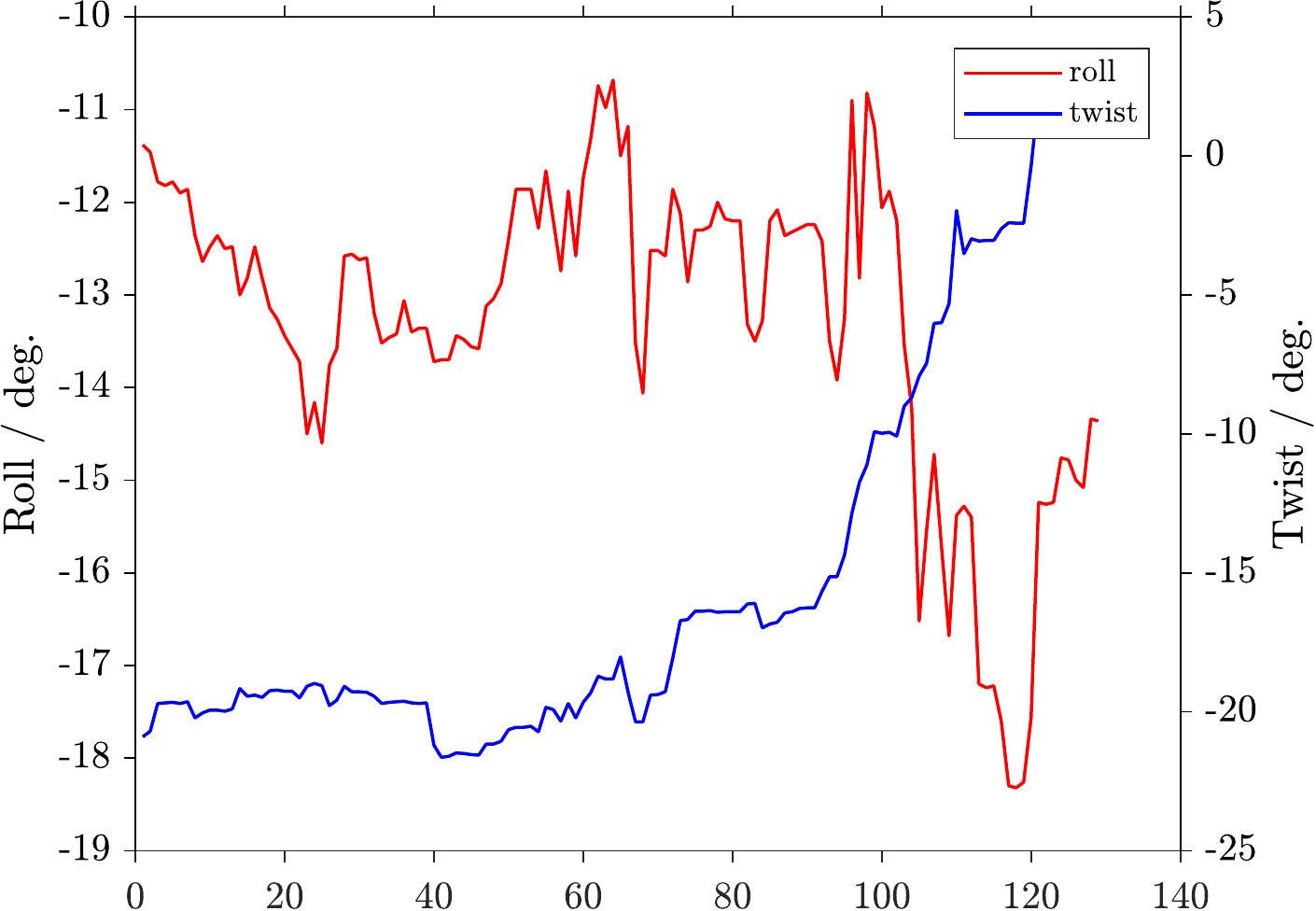}
\label{fig:62}
\end{subfigure}
\begin{subfigure}[b]{0.4\textwidth}
\includegraphics[width=\textwidth]{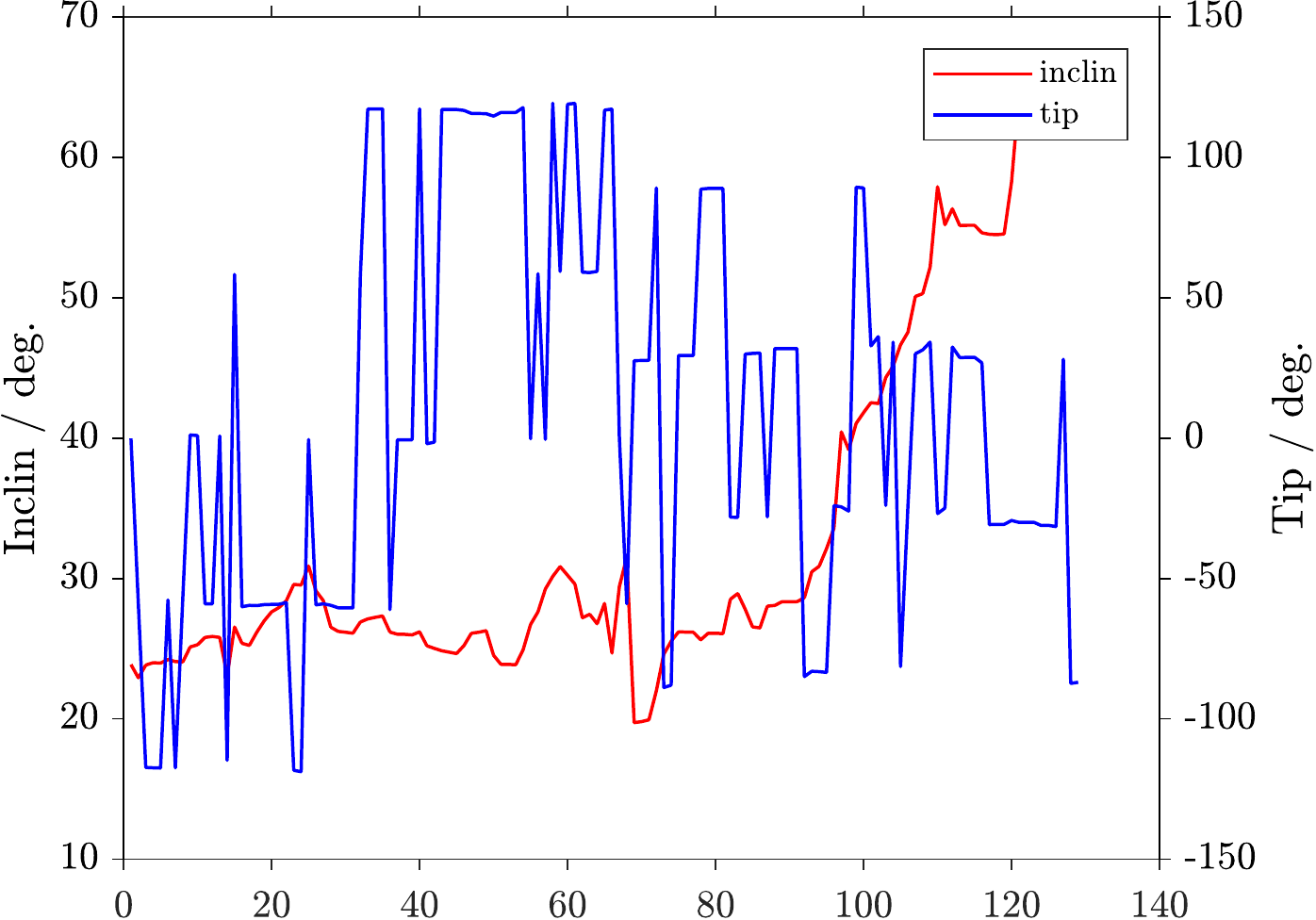}
\label{fig:63}
\end{subfigure}
\begin{subfigure}[b]{0.4\textwidth}
\includegraphics[width=\textwidth]{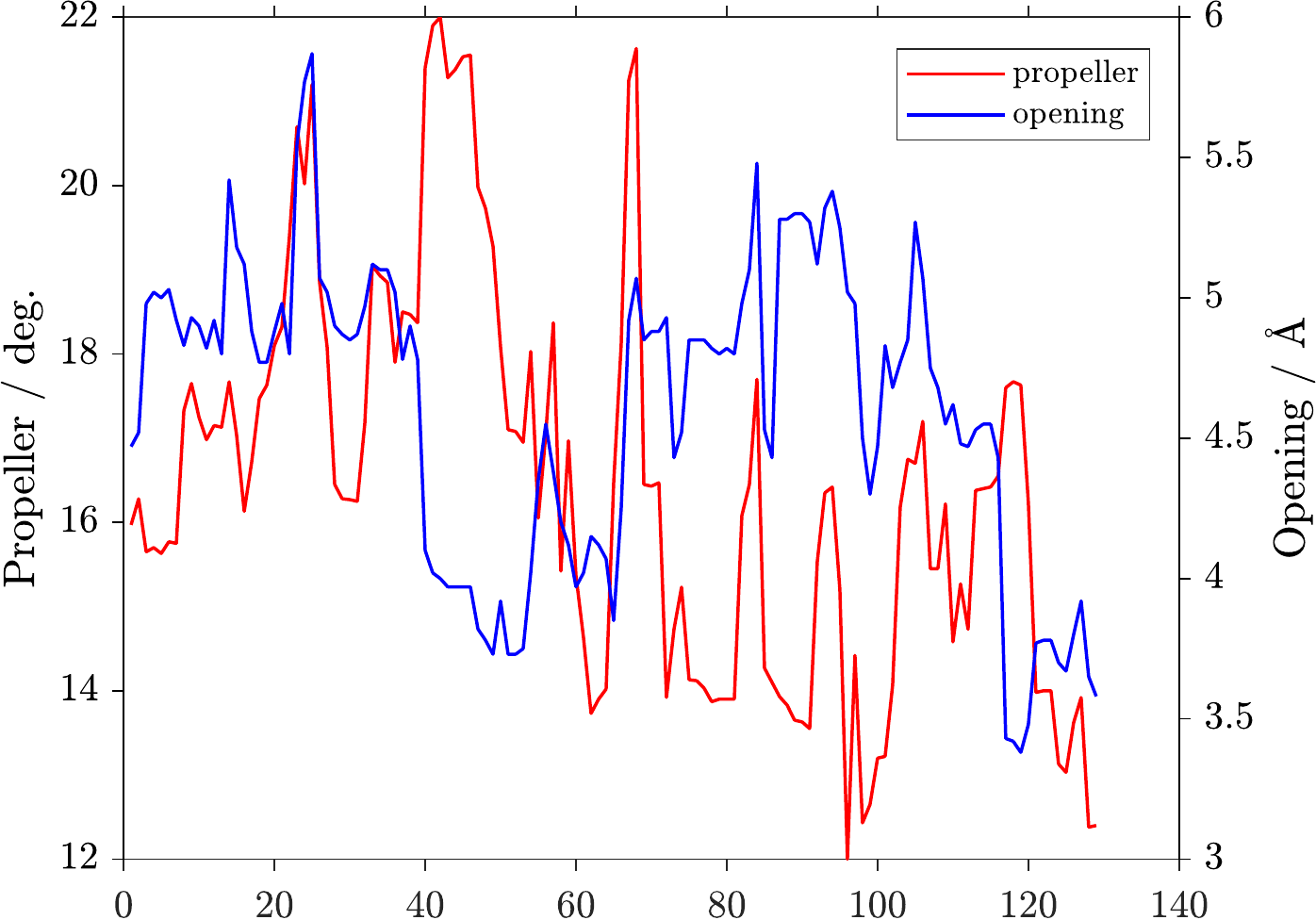}
\label{fig:64}
\end{subfigure}
\begin{subfigure}[b]{0.4\textwidth}
\includegraphics[width=\textwidth]{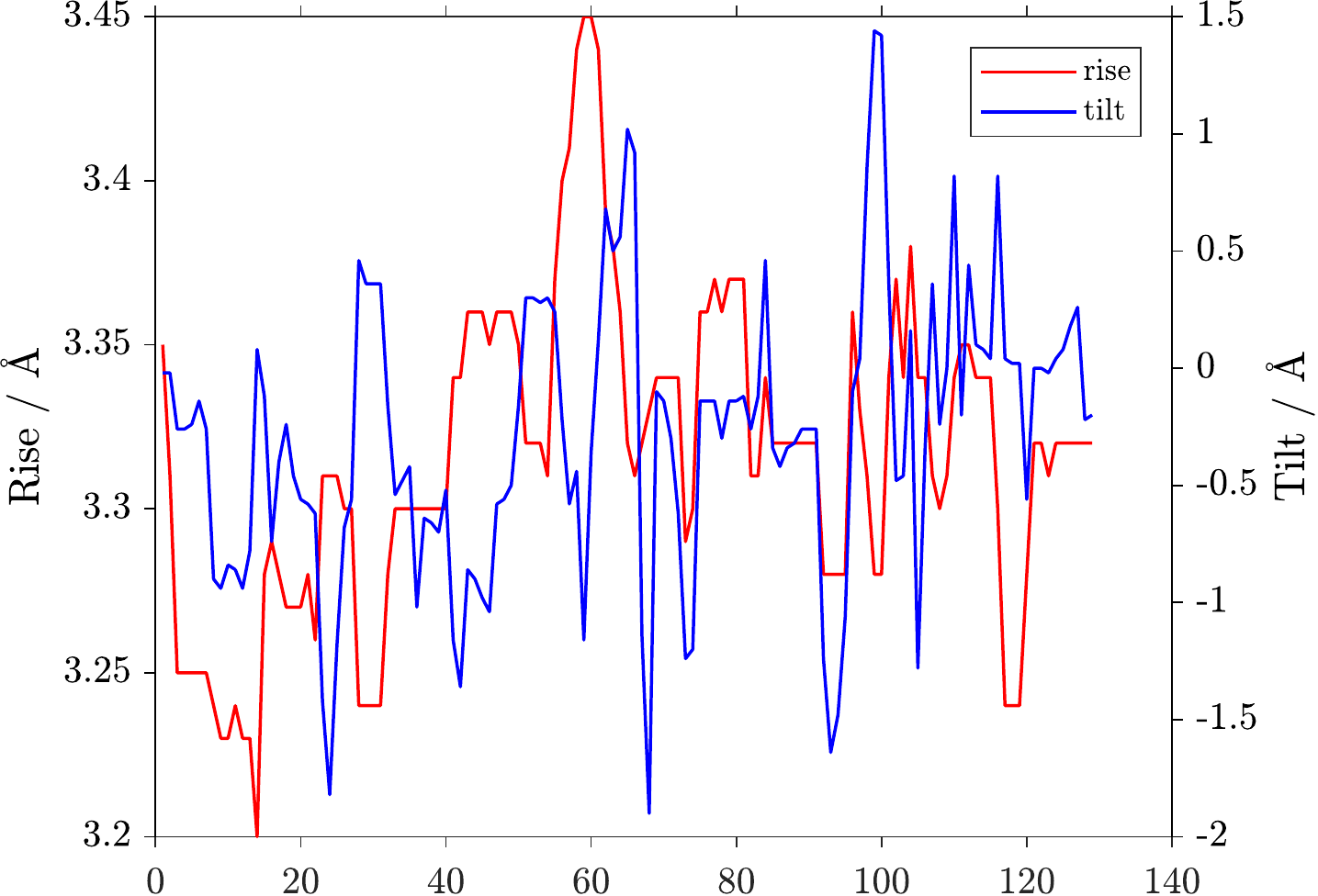}
\label{fig:65}
\end{subfigure}
\begin{subfigure}[b]{0.4\textwidth}
\includegraphics[width=\textwidth]{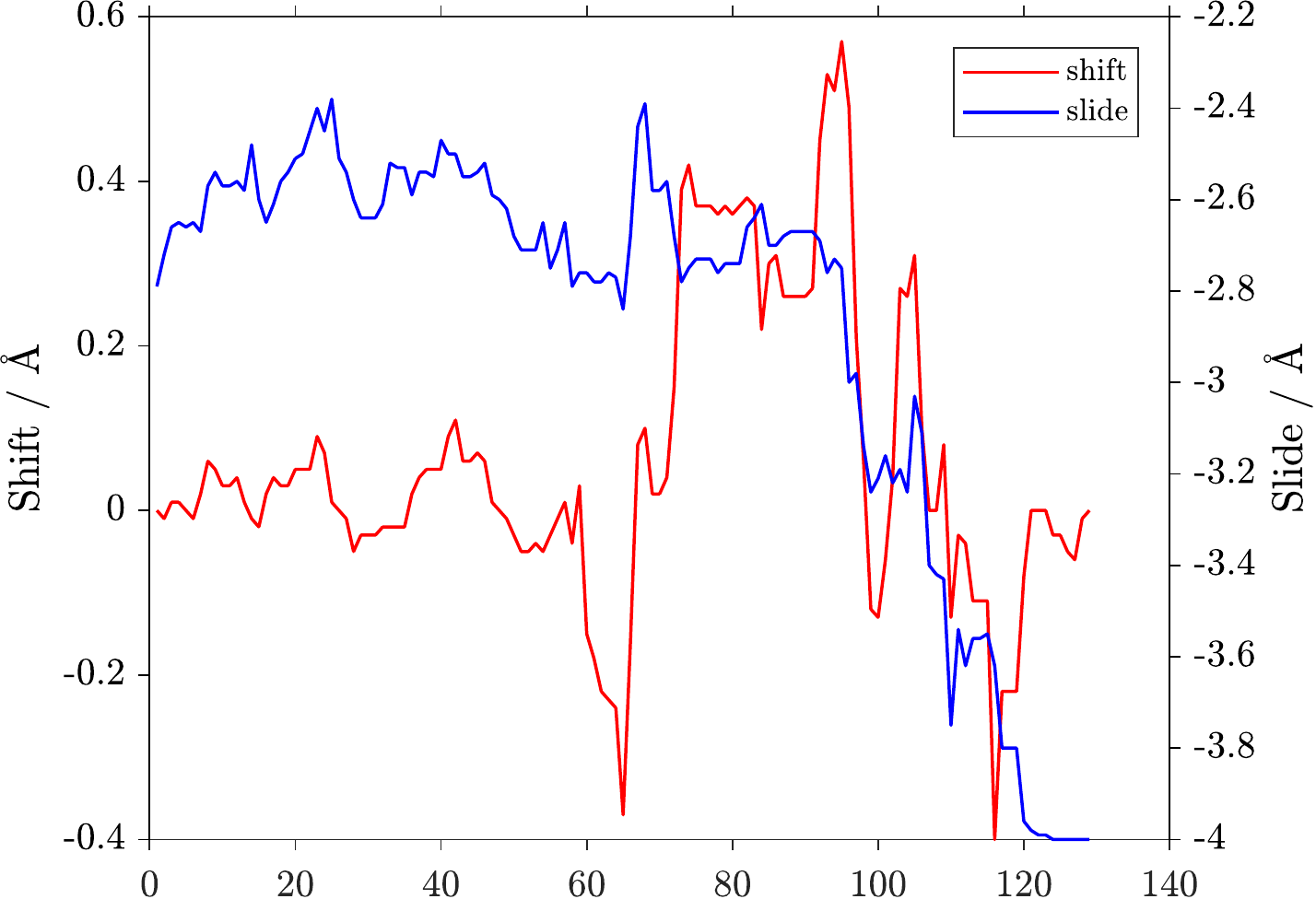}
\label{fig:66}
\end{subfigure}
\begin{subfigure}[b]{0.4\textwidth}
\includegraphics[width=\textwidth]{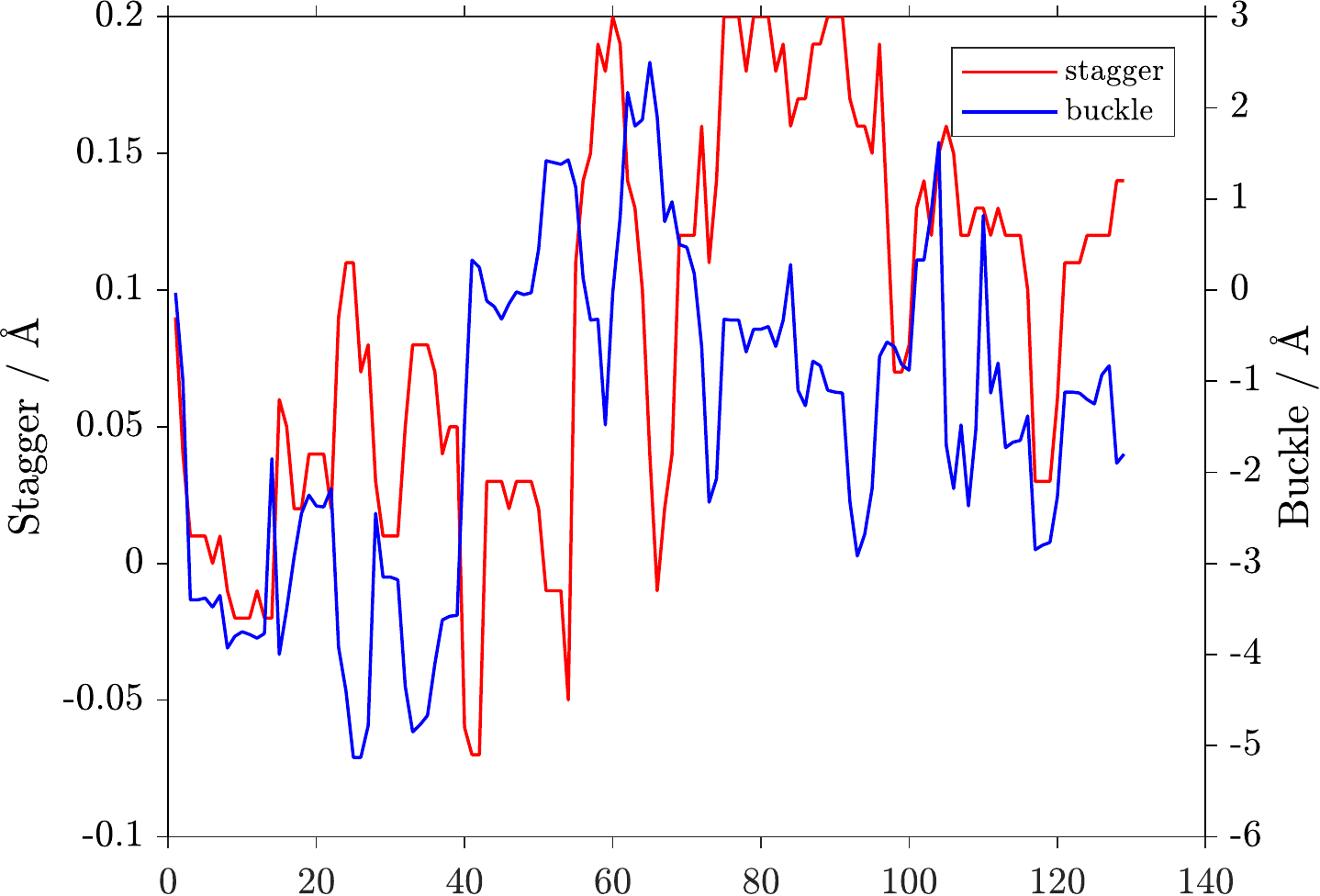}
\label{fig:67}
\end{subfigure}
\begin{subfigure}[b]{0.4\textwidth}
\includegraphics[width=\textwidth]{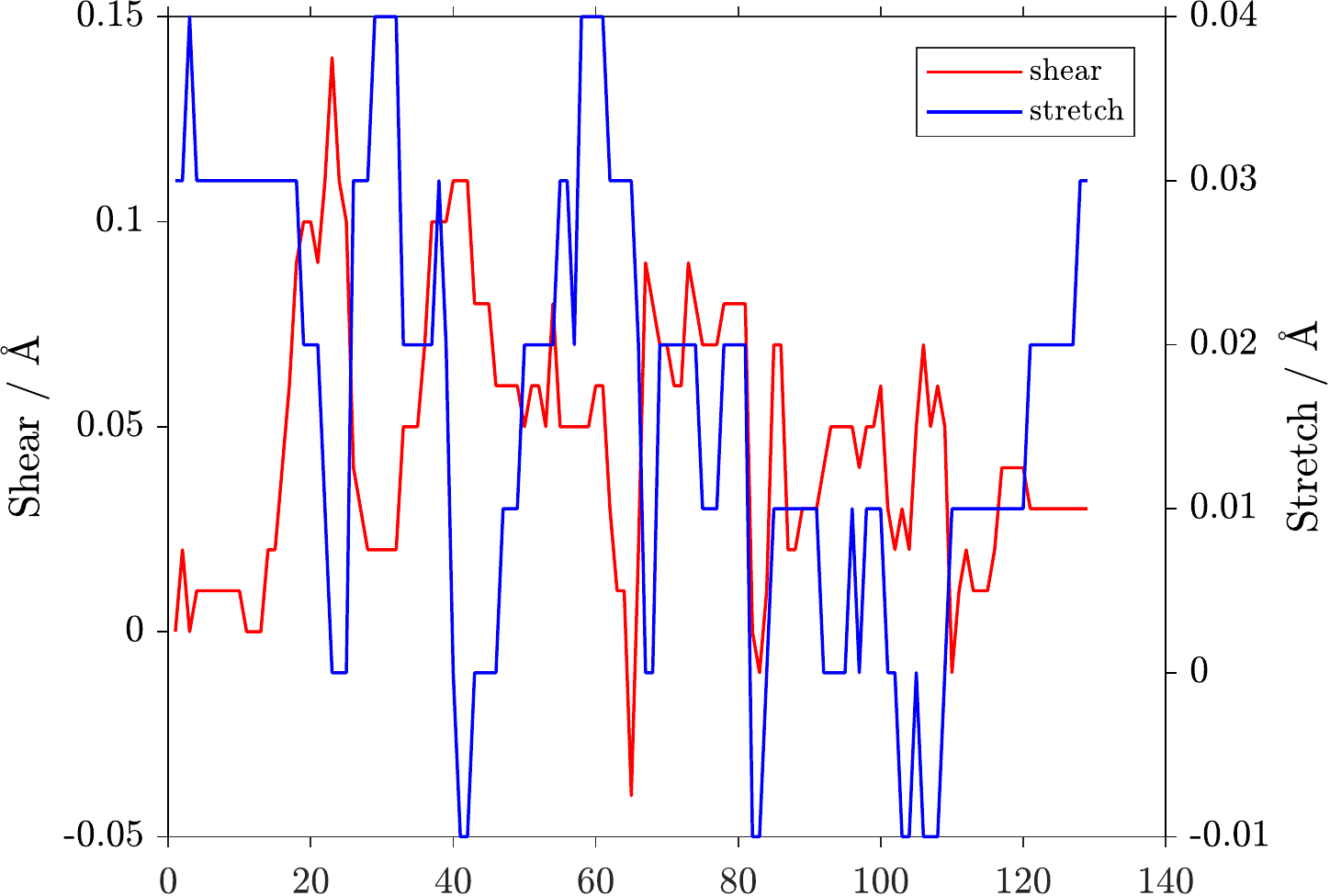}
\label{fig:68}
\end{subfigure}
\caption{Evolution of the bp-axis, inter-bp and intra-bp parameters along the fastest potential energy pathway for the right- to left-handed helical transition of the XyNA1 duplex, as a function of the position of the minimum in the discrete path.}
\label{fig:6}
\end{figure}

The evolution of backbone dihedral angles and the glycosidic torsion angle along the helical inversion pathways based on potential energy barriers is shown in Fig.\,\ref{fig:5}. The $\delta$ dihedral, with characteristic value \textit{ca}. $-40$\textdegree\enspace that effectively defines XyNA and dXyNA duplexes with respect to their natural analogues, remains relatively stable throughout the pathways (note the small scale). The $\alpha$, $\beta$ and $\gamma$ backbone dihedrals for the right-handed helical states of XyNA1 and dXyNA1 adopt values similar to those observed in canonical A-RNA and B-DNA, and undergo sign inversion in the course of the transition to left-handed helical states, seeded by the sign inversion of the $\delta$ dihedral with respect to the natural nucleic acid analogues. For the $\alpha$ and $\beta$ dihedrals, values in the left- and right-handed helical states are of approximately equal magnitude but opposite sign, in both XyNA and dXyNA. The behaviour of the $\epsilon$ and $\zeta$ dihedrals exhibits less variance along the pathways, though these angles do likewise adopt values of opposite sign to the corresponding angles in canonical A-RNA and B-DNA. The glycosidic torsion angle $\chi$ takes one of two predominant values in XyNAs, \textit{ca}. $-160$\textdegree\enspace (\textit{anti}) or \textit{ca}. $-80$\textdegree\enspace (\textit{syn}), and plays an important role in driving the transition. The evolution of inter-bp, intra-bp and bp-axis geometrical parameters\cite{luolsonjmolbiol1999} along the fastest potential energy pathway for the helical transition of XyNA1 is shown in Fig.\,\ref{fig:6}. There are large-scale changes in the bp-axis inclination angle, and in the communicative parameters of inter-bp twist angle, roll angle and slide distance. Values of the helical rise and helical twist are also diagnostic of each of the three major conformations.


\section{Conclusions}

The present results indicate that an equilibrium between left-handed helical and ladder-type structures, the conformational transitions between which are facile, exists for both XyNA and dXyNA duplexes. The global free energy minimum left-handed helical structure is more stable with respect to the ladder-type structure for the deoxyxylose analogue, one possible explanation for which is the increased solvation of the C2$'$ hydroxyl group, which is present in XyNAs but not dXyNAs, in the ladder-type compared to the helical state. The pair of conformational ensembles are more clearly well-defined for the dXyNA compared to the XyNA system, both with respect to the magnitude of the free energy barrier separating the basins and with respect to the helical handedness order parameter. Therefore the free energy landscapes demonstrate that XyNA duplexes are more flexible than dXyNA duplexes, and hence that the latter appear more promising candidates for use as a molecular switch, and for use as a chemical information storage molecule capable of self-replication, where facile unwinding of a helical structure is undesirable.

The free energy landscapes of both XyNA and dXyNA duplexes are significantly frustrated, highlighting an important factor that may have led evolution to select ribofuranosyl nucleic acids, and not the xylose-based analogues, as the genetic basis for life. The origin of the structural competition evident in these systems is the geometrical frustration that prevents Watson-Crick base pairing without the induction of strain in the nucleic acid backbone that must be relieved by bending, or else by the adoption of noncanonical base pairing modes at one or both of the duplex termini.

Free energy pathways from a disfavoured right-handed helical state to a left-handed helical state that is the global free energy minimum also differ notably between XyNA and dXyNA duplexes. For the latter system, extended linear structures represent an early and high-free energy transition state along the pathway, which then continues to proceed via regular helical structures with a smooth change in the handedness order parameter. For the XyNA system, the transition is mediated by low-energy ladder-type structures, which evolve to left-handed helical structures via a transition state ensemble of irregular `kinked' structures. Left-handed helix winding and unwinding transitions of XyNA and dXyNA duplexes are driven by the highly flexible terminal base pairs. The inversion of the $\delta$ dihedral angle in XyNAs with respect to their natural nucleic acid analogues seeds a direct inversion not only in the overall helical sense but also in the backbone dihedral angles. The glycosidic torsion angle also goes large-scale changes in the course of the transition, as do certain key geometric parameters, most notably the bp-axis inclination angle.

Further studies should focus on the design and application of XyNA and dXyNA duplexes for molecular devices, for example by investigating the sequence and length-dependence of the propensity for helicity, and the response of the equilibrium to environmental conditions. Kinetic studies could be used to compare the barriers for the helical unwinding transition of XyNA and dXyNA duplexes.

\begin{acknowledgement}

DJS gratefully acknowledges the Cambridge Commonwealth, European and International Trust for a PhD scholarship. KR and DJW gratefully acknowledge funding from the Engineering and Physical Sciences Research Council.

\end{acknowledgement}





\bibliography{xyna-bib}

\end{document}